\documentclass[10pt,letterpaper]{article}
\usepackage[top=0.85in,left=2.75in,footskip=0.75in,marginparwidth=2in]{geometry}

\usepackage[utf8]{inputenc}

\usepackage{cite}

\usepackage{nameref,hyperref}

\usepackage[right]{lineno}

\usepackage{makecell}
\usepackage{textcomp}
\usepackage{multirow}
\usepackage{amsthm,amsmath}
\usepackage{adjustbox}
\usepackage{booktabs}

\usepackage{microtype}
\DisableLigatures[f]{encoding = *, family = * }

\raggedright
\usepackage{changepage}

\usepackage[aboveskip=1pt,labelfont=bf,labelsep=period,singlelinecheck=off]{caption}

\makeatletter
\renewcommand{\@biblabel}[1]{\quad#1.}
\makeatother

\usepackage{lastpage,fancyhdr,graphicx}
\usepackage{epstopdf}
\fancyheadoffset[L]{2.25in}
\fancyfootoffset[L]{2.25in}

\usepackage{color}

\definecolor{Gray}{gray}{.25}

\usepackage{graphicx}

\usepackage{sidecap}

\usepackage{wrapfig}
\usepackage[pscoord]{eso-pic}
\usepackage[fulladjust]{marginnote}
\reversemarginpar

\begin{document}
\vspace*{0.35in}

% title goes here:
\begin{flushleft}
{\Large
\textbf\newline{Diagnostic Accuracy of Computed Tomography for Identifying Hospitalization in Patients with Suspected COVID-19 }
}
\newline
% authors go here:
\\
Morozov S.P.\textsuperscript{1},
Reshetnikov R.V.\textsuperscript{1, 2, *},
Gombolevskiy V.A.\textsuperscript{1},
Ledikhova N.V.\textsuperscript{1},
Blokhin I.A.\textsuperscript{1},
Kljashtorny V.G.\textsuperscript{1},
Mokienko O.A.\textsuperscript{1},
Vladzymyrskyy A.V.\textsuperscript{1}
\\
\bigskip
\bf{1} Research and Practical Clinical Center for Diagnostics and Telemedicine Technologies of the Moscow Health Care Department, Srednyaya Kalitnikovskaya str., 28, 109029, Moscow, Russia
\\
\bf{2} Institute of Molecular Medicine, Sechenov First Moscow State Medical University, Trubetskaya str. 8-2, 119991, Moscow, Russia
\\
\bigskip
* Corresponding author: reshetnikov@fbb.msu.ru

\end{flushleft}

\section*{Abstract}
\textit{Background and Objectives}
\newline
The use of computed tomography (CT) in COVID-19 screening is controversial. The controversy is associated with ambiguous characteristics of chest CT as a diagnostic test. The reported values of CT sensitivity and especially specificity calculated using reverse transcription polymerase chain reaction as a reference standard vary widely, raising reasonable doubts about the applicability of the method. The objective of this study was to reevaluate the diagnostic and prognostic value of CT using an alternative approach.
\newline
\textit{Methods}
\newline
This study included 973 symptomatic COVID-19 patients aged 42 $\pm$ 17 years, 56\% females. For all of them, we reviewed the disease dynamics between the initial and follow-up CT studies using a ``CT0-4" visual semi-quantitative grading system to assess the severity of the disease. Sensitivity and specificity were calculated as conditional probabilities that a patient's condition would improve or deteriorate, depending on the results of the initial CT examination. For the calculation of negative (NPV) and positive (PPV) predictive values, we estimated the COVID-19 prevalence in Moscow. The data on total cases of COVID-19 from March 6, 2020, to July 20, 2020, were taken from the Rospotrebnadzor website. We used several ARIMA and EST models with different parameters to fit the data and forecast the incidence.
\newline
\textit{Results}
\newline
The ``CT0-4" grading scale demonstrated low sensitivity (28\%), but high specificity (95\%). The best statistical model for describing the epidemiological situation in Moscow was ETS with multiplicative trend, error, and season type. According to our calculations, with the predicted prevalence of 2.1\%, the values of NPV and PPV would be 98\% and 10\%, correspondingly.
\newline
\textit{Discussion}
\newline
We associate the low sensitivity and PPV values of the ``CT0-4" grading scale with the small sample size of the patients with severe symptoms and non-optimal methodological setup for measuring these specific characteristics. We found that the grading scale was highly specific and predictive for identifying admissions to hospitals of COVID-19 patients. Despite the ambiguous accuracy, chest CT proved to be an effective practical tool for patient management during the pandemic, provided that the necessary infrastructure and human resources are available.

%\linenumbers

% the * after section prevents numbering
\section*{Introduction}
As of July 3, 2020, the COVID-19 pandemic led to more than 11 million confirmed cases globally, with nearly 525,000 deaths \cite{worldometers}. COVID-19 has a range of disease presentations. It may be completely asymptomatic or manifest itself with mild flu-like symptoms (80\% of infections), and there are severe and critical conditions requiring oxygen or ventilation (15\% and 5\%, correspondingly) \cite{WHO_report}. It is essential to control the disease spreading, but the variability of the symptoms makes the task of diagnosing COVID-19 very challenging. The availability of rapid and accurate COVID-19 testing tools becomes a critical factor in reducing human-to-human SARS-CoV-2 transition.

Currently, commercially available COVID-19 tests fall into two major categories: (i) assays detecting viral RNA via polymerase chain reaction (PCR) or nucleic acid hybridization, and (ii) serological and immunological assays detecting antibodies produced by individuals \cite{tests2020}. Despite constant improvement and evolving, both categories have their disadvantages. For example, reverse transcription PCR (RT-PCR) requires expensive equipment, highly trained operators, and can take days to provide results. As for the serological tests, there is currently no strong evidence of seropositivity correlation with immunity to the virus \cite{tests2020}. There remains an urgent demand for an effective tool for COVID-19 diagnosis, and computed tomography (CT) provides a promising solution.

CT is sensitive for early parenchymal lung disease, disease progression, and alternative diagnoses \cite{Fleischner}. Moreover, CT results can be obtained within 20 minutes, which is an essential factor during the pandemic. Nevertheless, there is considerable controversy regarding the use of CT imaging in COVID-19 diagnostics. The estimates of the CT chest sensitivity and sensibility vary widely and strongly depend on the radiologist’s competence (Table \ref{tbl:s1}).

RT-PCR tests, widely used as a reference standard in COVID-19 CT studies, also have limited diagnostic value. The probability of false-negative RT-PCR result decreases from 100\% on day 1 of infection to 68\% on day 4. On day 5 (typical time of symptom onset), the false-negative rate is 38\%, decreasing to 20\% on day 8 with the following increase to 66\% on day 21 \cite{RT_PCR_FN}. The sensitivity of RT-PCR tests might be as low as 59\% \cite{sens3} and depends on several factors, including individual variability in viral shedding \cite{pmid32277759}. It makes the method a sub-optimal reference standard. The objective of this study was to reevaluate the diagnostic value of CT through repeated scanning of patients with suspected COVID-19 that were either disposed to home care or hospitalized.
We used a recently reported CT grading scale \cite{CTxie} modified by us \cite{morozov2020} to assess the presence and severity of the disease. According to our results based on retrospective observations of 973 symptomatic subjects, CT has sensitivity 28\%, specificity 95\%, positive predictive value (PPV) 10\%, and negative predictive value (NPV) 98\% for identifying admissions to the hospitals of COVID-19 patients. Despite ambiguous characteristics, CT proved to be an effective practical tool for patient triage during the pandemic, which can significantly reduce the burden on hospitals.

\section*{Materials and Methods}

In this retrospective study, we analyzed primary and secondary CT results of individuals aged from 18 to 80 years examined at the Moscow Outpatient Computed Tomography Centers (OCTC) from April 4, 2020, to May 18, 2020. The patients eligible for inclusion were the subjects with the symptoms of acute respiratory infection (ARI). The clinical diagnosis of COVID-19 (ICD10 code: U07.2) was based on a combination of the ARI symptoms and CT results. CT images were acquired with recommended scanning parameters for standard-size patients (height, 170 cm; weight, 70 kg): voltage 120 kV, automatic tube current modulation, field of view 350 mm, slice thickness $\leq$ 1.5 mm. The CT images were inspected by the radiologists that completed the targeted training courses on COVID-19 chest imaging. In doubtful cases, the expert opinion of experienced radiologists was requested at the Moscow Radiology Reference Center.

CT results were categorized by the degree of lung tissue damage using a standardized template according to the grading scale ``CT0-4" \cite{morozov2020}. The patients from the categories CT0, CT1, and CT2, were disposed to home care under telemedicine guidance. The immediate hospitalization was mandatory for patients of the CT3 category. The CT4 category patients were referred to emergency medical care.

For statistical analysis, we examined whether the CT0-4 category may be associated with the prognosis for the disease progression, and therefore, the necessity of hospitalization. For that, we combined the CT results into two groups, the first of which, named ``home," contained patients of categories CT0-2 (home care), and the second, named ``hospital," included patients of categories CT3 and CT4 (hospital care). We considered the dynamics of the study participants between the two groups according to the results of secondary CT scanning to estimate the specificity and sensitivity of the grading scale. In our model, sensitivity was described as a conditional probability $P_{worse|hospital}$:

\begin{equation}
    P_{worse|hospital}=\frac{number \: of \:``hospital" \:patients \:after \:second \:CT \:study}{total \:number \:of \:``hospital" \:patients}
    \label{eqn:sens}
\end{equation}

The specificity of the model corresponded to a conditional probability $P_{better|home}$:

\begin{equation}
    P_{better|home}=\frac{number \:of \:``home" \:patients \:after \:second \:CT \:study}{total \:number \:of \:``home" \:patients}
    \label{eqn:spec}
\end{equation}

Positive predictive value (PPV) and negative predictive value (NPV) of a test are affected by the prevalence of the disease. We used Exponential Smoothing (ETS \cite{ETS}) and Auto-Regressive Integrated Moving Average (ARIMA \cite{ARIMA}) models to predict the COVID-19 incidence in Moscow. The daily data on the total number of COVID-19 cases in the city, covering the period from March 6, 2020, to July 20, 2020, was taken from the Rospotrebnadzor website \cite{RPN}. The time series analysis was performed with R 3.6.3 \cite{R} using forecast \cite{forecast} and ggplot2 \cite{ggplot} packages. For the accuracy of model estimation, we trained the model on the incidence data from March 6, 2020, to June 28, 2020, and compared predicted values with the actual values for a period from June 29, 2020, to July 20, 2020, using the mean absolute percentage error (MAPE) and mean absolute scaled error (MASE) metrics.

Using the prevalence value, PPV was calculated as follows:

\begin{equation}
    PPV=\frac{sensitivity*prevalence}{(sensitivity*prevalence) + (1-specificity)*(1-prevalence)}
    \label{eqn:ppv}
\end{equation}

Similarly, NPV of the test was defined as:

\begin{equation}
    NPV=\frac{specificity*(1-prevalence)}{specificity*(1-prevalence) + (1-sensitivity)*prevalence}
    \label{eqn:npv}
\end{equation}

\section*{Results}

\subsection*{Forecasting the COVID-19 prevalence in Moscow}

For choosing the forecast model, we separated the COVID-19 incidence data into training and testing portions and then applied different EST and ARIMA models to the training data. According to the MAPE and MASE values, ARIMA(0,2,1), ETS ZZZ (automatically selected parameters), and ETS ANN (simple exponential smoothing with additive errors) models provided the best fit for the testing dataset (Table \ref{tbl:2}).

\begin{table}[h!]
\caption{Accuracy statistics for different forecasting models}
\begin{adjustbox}{width=\textwidth}
    \begin{tabular}{p{4.4cm}p{4.4cm}p{4.4cm}}
    \toprule
    Model & MAPE & MASE \\
    \midrule
    ARIMA(0,2,1) & 0.56 & 0.66 \\
    ETS ZZZ & 0.56 & 0.66 \\
    ETS MMM & 1.08 & 1.28 \\
    ETS ANN & 0.56 & 0.66 \\
    \bottomrule
    \end{tabular}
\end{adjustbox}
\label{tbl:2}
\end{table}

\begin{wrapfigure}[21]{l}{83mm}
\includegraphics[width=82mm]{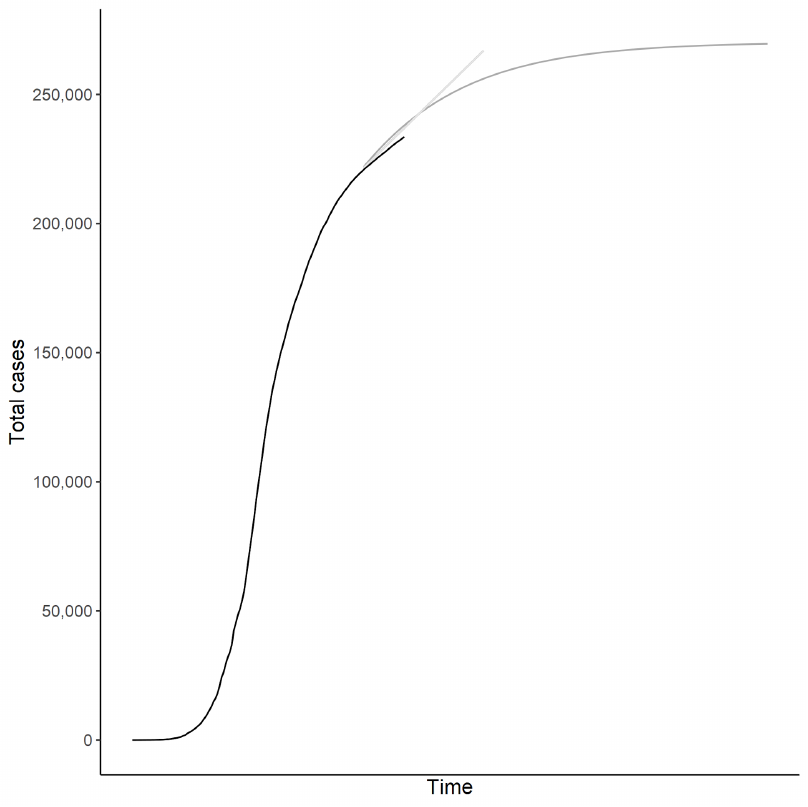}
\captionsetup{labelformat=empty}
\caption{}
\label{fgr:1}
\end{wrapfigure}

Despite the good fit with the training data, ARIMA(0,2,1), ETS ZZZ, and ETS ANN model predicted almost linear accumulation of COVID-19 cases. In contrast, the actual data showed a gradual decrease in daily incidence (Figure \ref{fgr:1}). The only model that followed the trend was ETS MMM (multiplicative trend, error, and season type), and we used it to estimate the number of total COVID-19 cases in Moscow.

\marginpar{
\vspace{.7cm}
\color{Gray}
\textbf{Figure \ref{fgr:1}. Forecasts of Moscow COVID-19 prevalence.} Black: actual data; gray: ETS MMM model, light-gray: ARIMA(0,2,1) model. Other models matched the predictions of ARIMA(0,2,1), and are not shown for clarity.
}

According to the ETS MMM model, the COVID-19 incidence tended to 270,000 subjects. Note that the model provides a very rough forecast horizon; there might be a second wave of the disease and seasonal patterns that we can not reliably predict using the current data. The Federal State Statistics Service reported that, as of January 1, 2020, the Moscow population was 12,678,079 \cite{GKS}. From these data, the COVID-19 prevalence in Moscow, defined as the percentage with the disease in the population at risk, would be 2.1\%.

\hfill \break

\subsection*{Estimating the diagnostic accuracy of CT}

\begin{figure*}[t!]
    \centering
    \includegraphics[scale=.77]{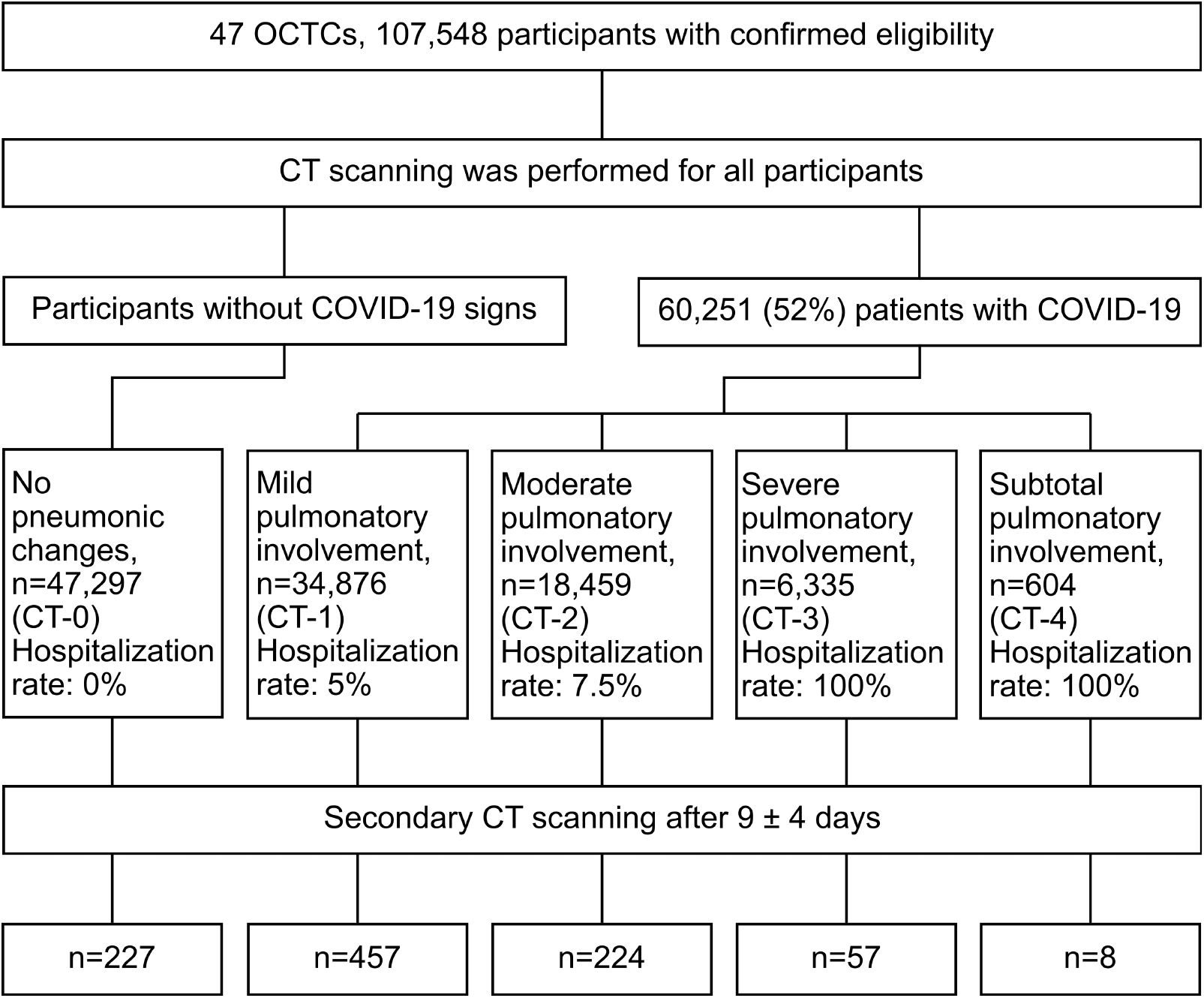}
    \caption{\textbf{Flow diagram of participants through the study.}}
\label{fgr:2}
\end{figure*}

47 OCTCs consulted 107,548 patients with a mean age of 42 $\pm$ 17 years, 60,539 females, assessed for initial eligibility (Figure \ref{fgr:2}). 47,297 (48\%) subjects showed no pneumonic changes (CT0 category). We collected the data of second CT examinations performed 9 $\pm$ 4 days after the first CT scan for a convenience sample of 973 patients from all five CT categories (Table \ref{tbl:3}).

\begin{table}[h!]
\caption{Categorization of participants between two consecutive examinations}
\begin{adjustbox}{width=\textwidth}
    \begin{tabular}{p{4.9cm}p{2.1cm}p{2.1cm}p{2.1cm}p{2.1cm}}
    \toprule
     &  & \multicolumn{2}{c}{Disease dynamics} &  \\
    \cmidrule{2-5}
    & & Better & Worse & Sum \\
    \cmidrule{2-5}
    \multirow{2}{*}{\parbox{4.9cm}{Group according to the initially assigned CT category}} & Home & 860 & 48 & 908 \\
    & Hospital & 47 & 18 & 65 \\
    \cmidrule{2-4}
    & Sum & 907 & 66 & \\
    \bottomrule
    \end{tabular}
\end{adjustbox}
\label{tbl:3}
\end{table}

We used data on the dynamics of the disease to assess the diagnostic and prognostic  value of the ``CT0-4" grading scale. For that, we considered the transition of a patient from the categories CT0-CT2 (``home" group) to CT3 or CT4 (``hospital" group) as deterioration of the condition (column ``Worse" in Table 3). The reverse transitions, from the ``hospital" group to the ``home" group, we regarded as an improvement in the patient's condition. Here we also included the cases in which the patient remained within the ``home" group, even when the disease progressed from CT0 to CT2 category (column ``Better" in Table 3). According to our calculations, the CT sensitivity was 27.7\%, specificity 94.7\%, PPV 10.1\%, and NPV 98.4\%.

\section*{Discussion}
The objective of this work was to estimate whether the CT method and ``CT0-4" grading scheme, in particular, have diagnostic and prognostic value as a test for COVID-19. Since RT-PCR tests, while being a ``gold standard" for COVID-19 clinical diagnosis, have several disadvantages \cite{RT_PCR_FN, sens3, pmid32277759}, we developed an alternative approach to assessing the characteristics of CT as a test for the disease. According to our results, the grading scheme demonstrates low sensitivity and PPV, but high specificity and NPV.

The role and value of CT in COVID-19 diagnostics have sparkled debates in the medical community \cite{pmid32402437, answer}. Chest CT has a low rate of missed COVID-19 diagnoses \cite{pmid32130038} and positively correlates with mortality in patients with COVID-19 pneumonia \cite{mortality}. On the other hand, CT does not test for the specific virus, and therefore its results could be misleading. Moreover, the experience and competence level of a radiologist plays an essential role in the correct recognition of COVID-19—specific CT patterns. Depending on the person interpreting the CT images, the specificity of the test could be as low as 7\% and as high as 100\% (see Table \ref{tbl:s1}).

When developing the ``CT0-4" grading system and the method of its implementation, we aimed to reduce the role and likelihood of human error in the interpretation of CT images. For this, we have introduced reference templates to assist the identification of both the presence and severity of the disease. Along with that, we established the MRRC, one of the tasks of which was to validate the decisions of radiologists at the OCTCs. Therefore, evaluating the diagnostic value of chest CT for COVID-19, we evaluate the efficiency of this entire structure.

Since CT is unable to distinguish between viruses and RT-PCR has several flaws that make it a sub-optimal reference standard, we monitored the clinical state for a convenience sample of 973 patients to address the correctness of the diagnosis. Note that the ``CT0-4" grading scale allows a radiologist to simultaneously make a diagnosis and a decision about the necessity of hospitalization, thus having predictive functionality. Because of that, we focused on the issue of whether the CT0-4 category provides a reliable prognosis for the disease progression in order to assess the diagnostic value of the approach.

Our results show that the ``CT0-4" grading system is highly specific and predictive for identifying admissions of COVID-19 patients to hospitals. Out of 908 patients initially assigned to the categories CT0-CT2, only 48 (5.3\%) progressed to the CT3 and CT4 categories, providing specificity 94.7\% and NPV 98.4\% for the approach. Of these 48 patients, only four (0.4\%) deteriorated to the CT4 category (two each from the initially assigned CT1 and CT2 categories). These numbers indicate the ability of the test to optimize the burden on hospitals due to reliable and efficient triage decisions.

We associate the low sensitivity and PPV values of the test with the features of the sample and the methodological setup. Due to the small number of patients in the ``hospital" group, even a slight change here can significantly affect the final value. For example, just one additional case in the ``Worse" column (see Table \ref{tbl:3}) would result in a sensitivity value being 29.3\% instead of 27.7\%. Moreover, we considered a rapid improvement of the patient's condition as a marker of a false-positive test result, although it could be a true-positive event. The median time to recovery for COVID-19 patients is estimated to be 20.8 days, but in individuals aged 50 \textendash 70+ years, it is longer (22.6 days), further lengthening for those with severe symptoms (28.3 days) \cite{pmid32353347}. In our study, the observation period was 9 $\pm$ 4 days, and subjects from the ``hospital" group were already at the advanced stages of the disease at the time of the initial CT examination. The improvement in the patient's condition during this period is consistent with the time to recovery values and therefore, may be associated with an effect of the therapy and not with the incorrect categorization of the individual. We consider it as a limitation of the study and recommend to treat the reported CT sensitivity and PPV values with caution and refer instead to the numbers in Table \ref{tbl:s1}.

During the pandemic, one of the essential properties of a diagnostic test is the ability to determine whether an asymptomatic subject has an undetected disease. For that, the test must (i) be capable of detecting the disease in its preclinical stage; (ii) be cost-effective; (iii) be widely available; (iv) be safe to administer, and (v) be able to demonstrate improved health outcomes \cite{ethics}. RT-PCR tests are safe and reasonable in cost, but their performance strongly depends on the stage of the disease, providing a non-constant false-negative rate \cite{RT_PCR_FN}.

Chest CT meets almost all the test requirements except safety. It is associated with potential harm from exposure to ionizing radiation, which makes it inappropriate for pregnant women and children. While CT is not widely available in general, some locations have the necessary infrastructure. For example, the network of Moscow city outpatient centers provided a sufficient amount of human and facility resources during the disease outbreak. A number of studies reported high sensitivity of CT \cite{sens1, sens2, sens3, sens4, sens5}. According to our results, the ``CT0-4" grading scale supported by the expert validation is able to compensate for the moderate sensibility of chest CT, making it a highly efficient tool for identifying admissions hospitals of COVID-19 patients.

\section*{Conclusions}
``CT0-4" grading scale demonstrated high sensibility for COVID-19 and was associated with the prognosis for the disease progression. These features make chest CT a rapid and effective diagnostic test for the disease during the pandemic, which simultaneously allows radiologists to make reliable triage decisions. The approach has its limitations related to the safety and availability of CT and human error factor in CT images interpretations. Substantial infrastructure capabilities, effective administrative decisions, and competent human resources are required to make its implementation appropriate. A network of OCTCs controlled by MRRC in combination with the ``CT0-4" grading scale proved to be an effective tool for routing and management of COVID-19 patients, which was able to optimize the burden on hospitals.

\section*{Acknowledgments}
The authors express their gratitude to all doctors of medical organizations of the Moscow Department of Health, fighting the epidemic, and to the team of experts from the Moscow Department of Information Technologies for prompt assistance in working with data from UMIAS-ERIS.

\nolinenumbers

%This is where your bibliography is generated. Make sure that your .bib file is actually called library.bib
\bibliography{library}

%This defines the bibliographies style. Search online for a list of available styles.
\bibliographystyle{ieeetr}

\clearpage

\section*{Supporting Information}

%These commands reset the figure counter and add "S" to the figure caption (e.g. "Figure S1"). This is in case you want to add actual figures and not just captions.
\setcounter{table}{0}
\renewcommand{\thetable}{S\arabic{table}}

\begin{table}[h!]
\caption{Sensitivity and specificity of CT for COVID-19 according to different sources}
\begin{adjustbox}{width=\textwidth}
    \begin{tabular}{p{2.1cm}p{2.1cm}p{9.3cm}}
    \toprule
        Sensitivity, \% & Specificity, \% & Comment \\
        \midrule
        98 & No data & 51 symptomatic patients, reference standard: RT-PCR  \cite{sens1} \\
        \cmidrule{1-2}
        93 & 53 & 103 symptomatic patients, reference standard: RT-PCR \cite{sens2}  \\
        \cmidrule{1-2}
        97 & 25 & 1014 symptomatic patients, reference standard RT-PCR \cite{sens3} \\
        \cmidrule{1-2}
        68 & 88 & 643 elective surgery and emergency patients, reference standard: RT-PCR \cite{sens4} \\
        \cmidrule{1-2}
        72 & 94 & \multirow{3}{*}{\parbox{9.3cm}{219 CT images of patients with confirmed COVID-19 pneumonia and 205 cases of non-COVID-19 pneumonia were blindly reviewed by three radiologists from China \cite{sens5} }} \\
        72 & 88 & \\
        94 & 24 & \\
        \cmidrule{1-2}
        80 & 100 & \multirow{7}{*}{\parbox{9.3cm}{An age-matched cohort of 58 randomly selected cases from the previous sample was blindly reviewed by four U.S. radiologists and three radiologists from China \cite{sens5} }} \\
        64 & 93 & \\
        97 & 7 & \\
        93 & 100 & \\
        83 & 93 & \\
        73 & 93 & \\
        70 & 100 & \\
         \bottomrule
    \end{tabular}
\end{adjustbox}
\label{tbl:s1}
\end{table}

\end{document}